\documentstyle[preprint,aps]{revtex}
\begin{document}
\pagestyle{plain}



\title{\bf PERMUTATION INVARIANT STATISTICS, DUALITY \\AND 
           SIMPLE INTERPOLATIONS }

\author{Bla\v zenka Meli\' c \thanks{e-mail: melic@thphys.irb.hr} and
        Stjepan Meljanac \thanks{e-mail: meljanac@thphys.irb.hr}}
\address{Institut Rugjer Bo\v skovi\'c, Bijeni\v cka 54, P.O.Box 1016, 10001 Zagreb, Croatia}
\date{\today}
\maketitle

\begin{abstract}
General permutation invariant statistics in the second quantized approach 
are considered.  
Simple interpolations between dual statistics are 
constructed. Particularly, we present a new minimal interpolation 
between parabosons and parafermions of any order. The connection with a simple 
mixing between bosons and fermions is established. The construction is 
extended to anyonic-like statistics. 
\end{abstract}
\vskip1.5cm

\def\ad{ a^{\dagger}}
\def\eea{ \end{eqnarray} }
\def\bea{ \begin{eqnarray} }
\thispagestyle{empty}
\newpage



%
%
In the last few years there has been increasing interest in generalized 
statistics. 
The main reason is their possible application to the theory of the fractional 
quantum Hall effect [1] and to the theory of anyon superconductivity [2] 
based on the two-dimensional concept of anyons.  
Haldane fractional statistics [3],
generalizing the Pauli exclusion principle to any spatial dimension, has also 
attracted much interest.
A large class of generalizations is based on permutation group invariance, 
for example parastatistics [4], 
 infinite quon statistics [5,6], 
 a simple interpolation between bosons and 
fermions introduced by Wu {\it et al.} [7] and Scipioni [8].
Similarly, braid group invariance leads to anyonic-like statistics. 
Recently, permutation invariant statistics has been studied in the first 
quantized approach [9],[10].

In this letter we follow the second quantized approach and present a 
unified view on all types of statistics invariant under the
permutation group. For a given type of generalized statistics we introduce 
the notion of
 its dual statistics and construct a simple interpolation between these two. 
Particularly, we analyse the minimal interpolation between 
Bose and Fermi statistics and some of its physical consequences, as well as the
minimal interpolation between para-Bose and para-Fermi statistics. 
We establish a connection with the statistics of Wu {\it et al.} [7] and 
Scipioni [8].  
Finally, we briefly discuss the extension of our simple interpolation to the 
anyonic-like statistics which are not permutation invariant.
\\
\\
{\bf Fock space and generalized statistics}

Let us consider a system of multi-mode oscillators described by $M$ pairs of 
creation and annihilation operators $a_{i}^{\dagger}$, $a_{i}$ ($i = 1,2,...,M$) hermitian conjugated to each other. We consider operator algebras with relations defined by a normally ordered expansion $\Gamma$ [11],
\begin{equation}
a_{i} a_{j}^{\dagger} = \Gamma_{ij}(a^{\dagger},a)
\end{equation}
and which possess 
the well-defined number operators $[N_{i},a_{j}^{\dagger}] = a_{i}^{\dagger}\delta_{ij}$,
\newpage 
\noindent
$[N_{i},a_{j}] = - a_{i}\delta_{ij}$ and $[N_{i},N_{j}] = 0$, $
i,j = 1,2,...,M$. In the associated Fock-like representation, let $|0\rangle$ 
denote the vacuum vector.\\ 
The scalar product is uniquely defined by $\langle 0|0\rangle = 1$, the vacuum 
condition $a_{i}|0\rangle = 0$, $a_i a_i^{\dagger}|0\rangle \neq 0$ and eq.(1). 
A general $N$-particle 
state is a linear combination of monomial state vectors 
$\ad_{i_{1}}\cdot\cdot\cdot 
\ad_{i_{N}}|0\rangle$, $i_{1},...,i_{N} = 1,2,...,M$.\\
We consider only relations (1) that may allow the norm zero vectors, but 
do not allow the state vectors of negative norm in the Fock space. 
 The norm zero vectors imply 
relations between the creation (annihilation) operators. These relations are 
consequences of eq.(1) and need not be postulated independently.
 
For a given $N$-particle monomial state $a^{\dagger}_{i_1}\cdot\cdot\cdot 
a^{\dagger}_{i_N}|0\rangle$ we write its type as $1^{n_1}
2^{n_{2}}...M^{n_{M}}$, where $n_{1},n_{2},...,n_{M}$ are multiplicities 
satisfying 
$n_{i}\geq 0$ and $\sum^{M}_{i=1} n_{i} = N$. There are in 
principle $N!/n_{1}!\cdot\cdot\cdot n_{M}!$ different states  
of the type $ 1^{n_1} 2^{n_{2}}...M^{n_{M}}$ and we define the corresponding 
matrix $A(n_{1},...,n_{M})$ of their scalar products. The number of linearly 
independent states is given by $d_{n_{1},...,n_{M}} = 
rank [A(n_{1},...n_{M})]$. The quantities $d_{n_{1},...,n_{M}}$ completely 
characterize the partition function and the thermodynamic properties of the 
free system defined by eq.(1).
Note that the partition function of the free system, i.e. the numbers 
$d_{n_{1},...,n_{M}}$, in general do not 
uniquely determine the operator algebra, eq.(1). 
The free Hamiltonian is defined by $H_{0} = \sum_{i=1}^{M} E_{i} N_{i}$, where 
$E_{i}$ and $N_{i}$ are the energy and the number operator corresponding 
to the $i^{\rm th}$ level. 
The partition function of the free system described by eq.(1) is given by
\begin{equation}
Z(x_{1},...,x_{M};\Gamma) = \sum^{\infty}_{N=0}\,\sum_{n_{1}+...n_{M}=N}\;
d_{n_{1},...,n_{M}}\, x_{1}^{n_{1}}\cdot\cdot\cdot x_{M}^{n_M}\,,
\end{equation}
where $d_{n_{1},...,n_{M}}$ is the degeneracy of the state 
with the energy $E = \sum_{i=1}^{M} E_{i}$  and $x_{i} = e^{-\beta/E_{i}},\,
 \beta = 1/kT$.\\
\\
{\bf Permutation invariant generalized statistics}

Our aim is to unify statistics [4-8] in the second quantized algebraic 
approach, eq.(1) by the simplest possible unifying principle with minimal
restrictions. It is permutation invariance, meaning that the matrix element
 $\langle 0|a_{i_{\pi (N)}} \cdot\cdot\cdot a_{i_{\pi (1)}}
\ad_{j_{\pi (1)}} \cdot\cdot\cdot \ad_{j_{\pi (N)}}|0\rangle$ does not 
depend on the permutation $\pi \in S_N$. 
Hence we assume that the set of relations defined by $\Gamma_{ij}$ in eq.(1) 
is 
invariant under the permutation group $S_{M}$. Then the coefficients in the 
expansion (1) do not depend on concrete indices in normal ordered monomials, 
but only on certain linearly independent types of permutation invariant terms, 
i.e.
\begin{equation}
a_{i}\ad_{j} = \delta_{ij} + C_{1,1} \;\ad_{j} a_{i} 
+ \sum_{n=1}^{\infty}\,\sum_{\pi,\sigma \in S_{n+1}}\;\;C_{\pi,\sigma}\,
\sum^{M}_{k_{1},...,k_{n} =1}\,[\pi(j,k_{1},...,k_{n})]^{\dagger}\,[\sigma(
i,k_{1},...,k_{n})]\,,
\end{equation}
where the operators $a_{i}$ are normalized in such a way that the 
coefficient of the $\delta_{ij}$ term is equal to $1$. 
The existence of the number operators $N_i$ implies that the annihilation 
and creation operators appearing in a monomial in the normal ordered expansion 
(3) have to come in pairs, i.e. monomials are diagonal in the variables
$k_1,...,k_n$ (up to permutations) [11]. 
The symbol $[\sigma(i,k_{1},...,k_{n})]$ 
denotes $a_{\sigma(i)}a_{\sigma(k_{1})}\cdot\cdot\cdot a_{\sigma(k_{n})} 
\equiv \sigma(a_{i}a_{k_{1}}\cdot\cdot\cdot a_{k_{n}})$. Also, 
$C_{\pi,\sigma} = C^{\ast}_{\sigma,\pi}$, owing to the hermiticity of the 
operator product $a_{i}\ad_{i}$. Furthermore, the $S_{M}$-invariant 
relations in eq.(3) acting on the 
corresponding Fock space imply the following relations:
\begin{equation}
a_{i}\ad_{i_{1}}\cdot\cdot\cdot\ad_{i_{N}}|0\rangle = \sum^{N}_{k=1}\,
\delta_{i i_{k}}\,\sum_{\sigma \in S_{N-1}}\,
\phi^{k}_{\sigma}[\sigma(i,..,\hat{i}_k,..,i_{N})^{\dagger}]\,|0\rangle\,,
\end{equation}
where $\hat{i}_{k}$ denotes the omission of the index $i_k$. 
The sum is running over all linearly independent monomials and 
$\phi^k_{\sigma}$ are 
(complex) coefficients. The identity $\phi^1_{id} = 1$ is implied by 
normalization in eq. (3). The coefficients $\phi^k_{\sigma}$ can be 
uniquely determined from $C_{\pi,\sigma}$ and vice versa.

The transition number operators $N_{i j}$, defined by the relations $[N_{i j},
\ad_{ k}] = \delta_{j k} \ad_{i}$ and  $N_{i i } \equiv N_{i}$, have a similar  expansion as $\Gamma_{i j}$ in eq.(3), namely
\begin{equation}
N_{i j} = \ad_{j} a_{i} + \sum^{\infty}_{n=1}\,\sum_{\pi,\sigma \in S_{n+1}}\; 
D_{\pi,\sigma}\,\sum^{M}_{k_{1},..,k_{n}=1} [\pi(j,k_{1},...,k_{n})]^{
\dagger} [ \sigma(i,k_{1},...,k_{n})]\,,
\end{equation}
where $D_{\pi,\sigma}$ are independent of $i,j$ (by 
permutation invariance) and $D_{\pi,\sigma} = D_{\sigma,\pi}^{\ast}$ following 
from $N_{i}^{\dagger} = N_{i}$. 
Hence, it follows that $N_{i j}^{\dagger} = N_{j i}$.

Each of the three sets of coefficients, $\{C_{\pi,\sigma}\}$, $\{
\phi^{k}_{\sigma}\}$, $\{ D_{\pi,\sigma}\}$, uniquely determines the two 
remaining 
sets, fixing the structure of the Fock space [11], and each of them is 
equivalent to the set of matrices $A(n_1,...,n_M)$.

The matrix $A(n_{1},...,n_{M})$ and its rank $d_{n_1,...,n_M}$ depend only 
on the collection of multiplicities $\{n_1,...,n_M\}$, which, written in
 the descending order $\lambda_1 \geq\lambda_2\geq ... \geq \lambda_M \geq 0, 
\,|\lambda|= \sum_{i=1}^{M}\lambda_i = N$, give rise to a partition 
$\lambda$ of $N$, 
i.e. $d_{n_1,...,n_M} = d_{\lambda}$ and $A(n_1,...,n_M) = A_{\lambda}$ [12]. 
If $\lambda_1 = \lambda_2 = \,...\,=\lambda_N = 1$, $\lambda_{N+1} = \,...
\,=\lambda_M = 0$, the corresponding Young tableau, denoted by $1^N$, is 
a column of $N$ boxes. The $N!\times N!$ generic matrix is denoted by 
$A_{1^N}$. All other matrices $
A_{\lambda},\, (|\lambda|=N)$ for any partition $\lambda$ of $N$ are easily 
obtained from the matrix $A_{1^N}$ [12,18]. 
The non-generic matix $A_{\lambda}$,
 for $\lambda \neq 1^N$, $\lambda_1 \geq \lambda_2 \geq ...
\lambda_k >0$, $|\lambda|=N$, is the matrix of the type $\frac{N!}{\lambda_1 !
\cdot\cdot\cdot\lambda_k !} \times \frac{N !}{\lambda_1 ! \cdot\cdot\cdot
\lambda_k !}$, whose matrix elements are enumerated by orbits $\bar{\alpha}$,
$\bar{\beta}$ of permutations $\alpha,
\beta \in S_N$ acting on the multi-set $\{i_1 \leq i_2 ... \leq i_N\}$
with multiplicities $\lambda_1,\lambda_2,...,\lambda_k$, 
and
\begin{displaymath}
(A_{\lambda})_{\bar{\alpha},\bar{\beta}} = \sum_{\sigma \in S_N,
 \sigma\bar{\beta}
 = \bar{\beta}} (A_{1^N})_{\alpha , \sigma\beta}.
\end{displaymath} 
(Note that the matrix elements do not depend on $i_1,...,i_N$, only on their 
multiplicities.)

By the permutation symmetry of $\Gamma$, 
eq.(3), it follows that $A_{1^N}$ can be written as
\begin{equation}
A_{1^N} = \sum_{\pi \in S_N} f(\pi) R(\pi)\,,
\end{equation}
where $R$, $R(\pi)_{\mu , \nu} = \delta_{\mu\pi , \nu}$ is the right regular 
representation of the permutation group $S_N$ 
and $f(\pi)$ are complex numbers completely determining all matrix elements 
and statistics. 
The matrix $A_{1^N}$ is hermitian with 
non-negative eigenvalues and rank $d_{1^N} \leq N!$.

The $S_M$-invariant partition function can be expanded into the form
\begin{equation}
Z_N(x_1,...,x_M) = \sum_{\lambda,|\lambda|=N}\;d_{\lambda} \,m_{\lambda}
(x_1,..., x_M) = \sum_{\mu,|\mu|=N}\;n(\mu) s_{\mu}(x_1,...,x_M)\,,
\end{equation}
where $ m_{\lambda}(x_1,...,x_M)$ is the monomial $S_M$-invariant function and 
$s_{\mu}(x_1,...,x_N)$ is the Schur function [13] satisfying
\begin{equation}
s_{\mu}(x_1,...,x_M) = \sum_{\mu}\,K_{\mu\lambda}\, m_{\lambda}(x_1,...,x_M)\,.
\end{equation}
Hence, from eqs.(7) and (8) it follows that
\begin{equation}
d_{\lambda} = \sum_{\mu}\,n(\mu) K_{\mu \lambda}\,,
\end{equation}
where $n(\mu)$, $dim (\mu) \geq n(\mu)\geq 0$, is the number of equivalent 
IRREP's (irreducible 
representation) $\mu$ of physical states contributing to the decomposition of 
$A_{1^N}$, and $K_{\mu \lambda}$ are Kostka's numbers 
denoting the number of linearly independent states \\
$\ad_{i_1}\cdot\cdot\cdot \ad_{i_N}|0\rangle $ of type 
$\lambda$ which fill the Young frame $\mu$ in the column strict way and
 $K_{\mu \lambda} \leq K_{\mu 1^N} = dim (\mu)$. The number of $N$-particle 
independent states is $D(M,N) = Z_N(1,1,...,1)$. 

The numbers 
$n(\mu)$ completely determine the partition function of 
the free $S_M$-invariant system defined by eq.(1), but do not determine 
the operator algebra itself. 
The two free systems with the same partition function can differ in the 
following properties: (i) in the commutation relations of their creation 
(annihilation) operators, (ii) in the probabilities of finding the monomial
state $\ad_{i_1}\cdot\cdot\cdot\ad_{i_N}|0\rangle$ in the IRREP $\mu$ of $S_N$, 
(iii) in the probabilities of finding the particular IRREP $\rho_k$ of $S_{n_1+
n_2}$ in the decomposition of $\mu_1 \times \mu_2 = \sum_k \rho_k$, where 
$\mu_i$ is the IRREP of $S_{n_i}, i = 1,2$ and (iv) in the probabilities of 
finding a particular subsystem characterized by the IRREP $\mu_1 \times \mu_2
\cdot\cdot\cdot$ of $S_{n_1}\times S_{n_2}\times\cdot\cdot\cdot$ in the 
larger system $\mu$ of $S_n, n \geq n_1+n_2+\cdot\cdot\cdot$.\\ 
 Examples of
permutation invariant statistics defined by eq.(1) are parastatistics [4], 
interpolation between parastatistics [14] and infinite quon statistics [5,6]. 
\\
\\
{\bf Duality and simple interpolation}

Let us first discuss a duality between Bose and Fermi statistics. For Bose 
statistics, $f(\pi) = 1$ in the expression (6), and for Fermi statistics, 
$f(\pi)=
(-)^{I(\pi)}$, $\forall\pi\in S_N$, where $I(\pi)$ is the number of inversions 
in $\pi$. Hence, the Bose and Fermi generic matrices are 
hermitian of rank one with the same spectrum (generic matrices are similar).
 If these properties of the Bose and Fermi generic matrices were
true for all partitions $\lambda$, Bose and Fermi statistics would be 
the same. However, for Bose statistics, all matrix elements are equal to 
$\lambda_1 !\cdot\cdot\cdot\lambda_k !$ (the rank is one with the eigenvalue 
$N !$) and for Fermi statistics, all matrix elements are zero. Hence, the 
crucial difference between Bose and Fermi statistics is in the structure of 
non-generic matrices. However, a
duality transformation between completely symmetric (Bose) and 
antisymmetric (Fermi) eigenvectors of generic matrices can be defined. 

Here we generalize this duality between Bose and Fermi statistics to any 
permutation 
invariant statistics defined by a set of generic matrices $A_{1^N}$. The dual
generic matrix $A^d_{1^N}$ is given by
\begin{equation}
A^d_{1^N} = D_{1^N}\, A_{1^N} \,D_{1^N}\,,
\end{equation}
where $D_{1^N}$ is the $N! \times N!$ diagonal matrix with matrix elements
\begin{equation}
(D_{1^N})_{\pi,\sigma} = (-)^{I(\pi)}\delta_{\pi,\sigma}\,,
\end{equation}
where $\pi$, $\sigma \in S_N$ and $I(\pi)$ is the number of inversions of 
permutation $\pi$. We point out that the duality trasformation has non-trivial 
consequences on the non-generic matrices $A_{\lambda}$, $\lambda \neq 1^N$.
\\ 
It follows that 
$D^{\dagger}_{1^N} = D_{1^N}\, ,\, D^2_{1^N} = 1$ and $Tr D_{1^N} = 0$. If 
$A_{1^N} = \sum_{\pi} f(\pi) R(\pi)$, eq.(6), then $A^d_{1^N} = \sum_{
\pi} f^d(\pi) R(\pi)$, where $f^d(\pi) = (-)^{I(\pi)} f(\pi)$ since 
$D_{1^N} R(\pi) D_{1^N} = (-)^{I(\pi)} R(\pi)$. Furthermore, we have 
the following proposition:\\
{\it Proposition.} If the matrix $A_{1^N}$ is hermitian, then $A^d_{1^N}$ is 
also 
hermitian and possesses the same eigenvectors and spectrum as 
$A_{1^N}$. Hence, $A^d_{1^N}$ and $A_{1^N}$ commute.\\
{\it Proof.} Let us denote the eigenvectors of $A_{1^N}$ as $|a,b,\mu\rangle$,
$a,b = 1,2,...,dim(\mu)$, where $\mu$ fixes the IRREP of $S_N$, $b$ enumerates 
equivalent IRREP's $\mu$ and $a$ enumerates states in the $b^{th}$ IRREP $\mu$, in accordance with the decomposition of a regular representation. 
The components of the given eigenvector $|a,b,\mu\rangle$ are
\begin{equation}
|a,b,\mu\rangle = \sum_{\pi\in S_N} R^{\mu}_{a,b} (\pi)|\pi\rangle\,,
\end{equation}
where $R^{\mu}$ is the unitary IRREP $\mu$ of $S_N$, characteristic of the 
generic matrix $A_{1^N}$.

The corresponding eigenvalue of $A_{1^N}$ is $\Lambda^{\mu}_b$
\begin{eqnarray}
A_{1^N}|a,b,\mu\rangle &=& \Lambda^{\mu}_{b}|a,b,\mu\rangle \, ,\; a=1,2,...,
dim(\mu)\,, \nonumber\\ 
\Lambda_{b}^{\mu} &=& \sum_{\pi\in S_N} R^{\mu}_{bb}(\pi)f(\pi)\,.
\end{eqnarray}
Let us show that $|a,b,\mu\rangle$ are the eigenvectors of $A_{1^N}$ as well:
\begin{eqnarray}
A^d_{1^N}|a,b,\mu\rangle &=& D_{1^N} A_{1^N} D_{1^N}|a,b,\mu\rangle  = 
D_{1^N}A_{1^N}|a',b',\mu^{T}\rangle \nonumber\\
&=& D_{1^N}\Lambda^{\mu^T}_{b'}|a',b',\mu^T\rangle = \Lambda^{\mu^T}_{b'}
|a,b,\mu\rangle\,, \nonumber\\
\Lambda^{\mu^T}_b &=& \sum_{\pi\in S_N} R^{\mu^T}_{bb}(\pi) f(\pi) = \sum_{
\pi\in S_N} (-)^{I(\pi)} R^{\mu}_{bb}(\pi) f(\pi)\,,
\end{eqnarray}
since $D_{1^N}|a,b,\mu\rangle = 
|a',b',\mu^T \rangle$ and $R^{\mu^T}(\pi) = (-)^{I(\pi)} R^{\mu}(\pi)$. 
Hence, the operator $D$ transforms the IRREP $\mu$ to its dual IRREP $\mu^T$. 
The
$A^d_{1^N}$ and $A_{1^N}$ have common eigenvectors $|a,b,\mu\rangle$ with 
eigenvalues $(\Lambda^d)^{\mu}_b = \Lambda^{\mu^T}_b$ and $\Lambda^{\mu}_b$, 
respectively.

We point out that the complete characterization of the statistics 
considered is not given 
only by $n(\mu)$'s (determining the number of equivalent IRREP's $\mu$ 
of independent physical states that 
contribute), but also requires the eigenvalues $\Lambda^{\mu}_b \geq 0, \forall 
\mu,b$,
which determine all relevant probabilities of finding the monomial state 
$\ad_{i_1}\cdot\cdot\cdot\ad_{i_N}|0\rangle$ in equivalent IRREP's $\mu$ of 
$S_N$, i.e. $w(\mu) = K_{\mu\,1^N}/N!\,\sum_b \Lambda^{\mu}_b$. Hence, 
although the dual generic matrix  $A^d_{1^N}$ has the same eigenvectors as 
$A_{1^N}$ and $d^d_{1^N} = d_{1^N}$, 
they basically differ since $\Lambda^{\mu^T}_b \neq \Lambda^{\mu}_b$.
Thus, the two permuation invariant statistics related through the duality 
transformation, 
eq.(10), are qualitatively different. They are connected by conjugation of 
the Young tableaux. 
\\
\\
{\bf Simple interpolation}

There is a general simple construction of mixing the given permutation 
invariant statistics with its dual statistics. Since our generic matrices 
$A_{1^N}$ 
have non-negative eigenvalues, their dual generic matrices have 
non-negative eigenvalues, too. The mixed generic matrices $A^q_{1^N}$ are 
defined by
\\
\begin{eqnarray}
A^q_{1^N} &=& \frac{1+q}{2} A_{1^N} + 
\frac{1-q}{2} A^d_{1^N} = \sum_{\pi \in S_N} f^q(\pi) R(\pi)\,, \\
f^q(\pi) &=& \left\{ \begin{array}{ll}
                 f(\pi) & \;\;{\rm for}\;\; \pi \;\;{\rm even}\\
                 q f(\pi) & \;\;{\rm for}\;\; \pi \;\;{\rm odd}
                 \end{array}
         \right. \,,\nonumber
\end{eqnarray}
with $|q| \leq 1$. The matrix $A^q_{1^N}$ has the same eigenvectors as 
$A_{1^N}$, 
with eigenvalues $\Lambda^q_{1^N}$:
\begin{eqnarray}
A^q_{1^N}|a,b,\mu\rangle &=& (\Lambda^q)^{\mu}_b|a,b,\mu\rangle \,,\nonumber\\ 
(\Lambda^q)^{\mu}_b &=& \frac{1+q}{2}\Lambda^{\mu}_b + \frac{1-q}{2}
\Lambda^{\mu^T}_b \geq 0,\; |q| \leq 1.
\end{eqnarray}
It is obvious that $(\Lambda^q)^{\mu}_b \geq 0$ if $\Lambda^{\mu}_b \geq 0$ and 
$|q|\leq 1$ and that $d_{1^N} \leq d^q_{1^N} \leq 2 d_{1^N}$. \\
Let us analyse a few examples.
\\
\\
{\bf Minimal interpolation between bosons and fermions}

Here we construct the minimal generalized statistics with permutation group 
invariance, interpolating between Bose and Fermi statistics. The generic 
matrix $A_{1^N}$, eq.(6), for Bose statistics is characterized by $f(\pi) = 1$,
 for all $\pi \in S_{N}$ and for Fermi statistics by 
$f(\pi) = 1$, if $\pi$ is an even permutation and $f(\pi) = -1$ if $\pi$ is 
 an odd permutation. We suggest a simple interpolation defined by
\begin{equation}
f^q(\pi) = \left\{ \begin{array}{ll}
                 1 & \;\;{\rm for}\;\; \pi \;\;{\rm even}\\
                 q & \;\;{\rm for}\;\; \pi \;\;{\rm odd}
                 \end{array}
         \right. \,,
\end{equation}
where $|q| \leq 1$. Note that $q = 1\,(-1)$ corresponds to Bose (Fermi) 
statistics. The non-zero non-degenerate eigenvalues of the matrix $A_{1^N}$ are
$(1+q)N!/2$ and $(1-q)N!/2$ with eigenvectors in the 
symmetric and antisymmetric representation, respectively. Hence, if $|q| < 1$,
the rank of the matrix $A_{1^N}$ is $d_{1^N} = 2$ and the Fock space does not
 contain vectors of negative squared norm. Then it follows that the multiplicity
 is $n(\mu) = 1$ for $\mu = 1^N$ and $\mu = N$, and $n(\mu) = 0$ otherwise. 
Null vectors imply that the state vectors are divided into two classes:
\begin{equation}
\begin{array}{ll}
\pi (\ad_1\cdot\cdot\cdot\ad_N) \equiv \ad_1\ad_2\cdot\cdot\cdot\ad_N|0
\rangle \, , \; \pi \;{\rm even}\,,\\
\pi (\ad_1\cdot\cdot\cdot\ad_N) \equiv \ad_2\ad_1\ad_3\cdot\cdot\ad_N|0
\rangle\, , \; \pi \;{\rm odd}\,, 
\end{array}
\end{equation}
and any generic monomial state can be decomposed into the sum of symmetric 
and antisymmetric states:
\begin{equation}
\ad_1\cdot\cdot\cdot\ad_N|0\rangle = \frac{1}{2}(\ad_1\ad_2 + \ad_2\ad_1)\,
\ad_3\cdot\cdot\cdot\ad_N|0\rangle + \frac{1}{2}(\ad_1\ad_2 - \ad_2\ad_1)\,
\ad_3\cdot\cdot\cdot\ad_N|0\rangle\,.
\end{equation}
The probability for the state $\ad_1\cdot\cdot\cdot\ad_N|0\rangle$ to be found 
in the symmetric (resp. antisymmetric) state is $w_s = (1+q)/2$ 
(resp. $w_a = (1-q)/2)$.\\
The matrix $A_{\lambda}$, $\lambda \neq 1^N$, has rank $d_{\lambda} = 1$ and it
is identical to the matrix $A_{\lambda}$ for Bose statistics, $A_{\lambda}
^B$, i.e completely has the Bose character,
\begin{equation}
A_{\lambda}^q = \frac{1+q}{2}\,A_{\lambda}^B \;.
\end{equation} 

Using the generic matrix $A_{1^N}$ one easily finds
\begin{equation}
a_i\ad_{i_1}\ad_{i_2}|0\rangle = [\,\delta_{i i_1} \ad_{i_2} + q 
\delta_{i i_2}\ad_{i_1}]\,,
\end{equation}
and for $N \geq 3$,
\bea
a_i\ad_{i_1}\cdot\cdot\cdot\ad_{i_N}|0\rangle = &[& \delta_{i i_1} 
(i_2,...,i_N)_{\rm even} +
\delta_{i i_2} (i_1,i_3,...,i_N)_{\rm odd} + 
\, ... \,\nonumber\\
&+& \delta_{i i_N} 
(i_1,i_2,...,i_{N-1})_{\rm even(odd)}\,]\,|0\rangle\,.
\eea
The subscript in the last term denotes even(odd) permutations for
$N$ odd(even). 

If all indices are equal, eqs.(21,22) imply
\begin{equation}
\begin{array}{ll}
a_i(\ad_i)^2|0\rangle = (1+q) \ad_i |0\rangle \,,\\
a_i(\ad_i)^n|0\rangle = n (\ad_i)^{n-1}|0\rangle\, ,\; n \neq 2\,.
\end{array}
\end{equation}
Then for a single oscillator [15], one can write 
$\ad a = \varphi (n)\, ,\; a\ad = \varphi (n+1)$ and 
$\langle 0|a^n(\ad)^n|0\rangle = [\varphi (n)]! = \frac{1}{2}(1+q)n!$, where
\begin{equation}
\varphi (n) = \left\{ 
             \begin{array}{ll}
              n & \;\; n \neq 2 \\
              1+q & \;\; n=2 \,.
            \end{array}
              \right.
\end{equation}

The expansion in eq.(3) implied by eqs.(21,22) can be written as 
follows:
\bea
a_i\ad_j = \delta_{ij} &+& q \ad_j a_i \nonumber\\
         &+& \sum^{M}_{k=1}\,[x_1 (jk)^{\dagger}(ik) + 
                   z_1(kj)^{\dagger}(ki) +
                   y_1 (kj)^{\dagger}(ik) + 
                   y_1 (jk)^{\dagger}(ki)] \nonumber\\
         &+& \sum_{n=2}^{\infty}\;\sum^{M}_{k_1,...,k_n=1}\;
                   [x_n(jk_1\cdot \cdot\cdot k_n)^{\dagger} + 
                   y_n(jk_1\cdot\cdot\cdot k_n k_{n-1})^{\dagger}]
                   (ik_1\cdot\cdot\cdot k_n)\,,
\eea
and for $|q| < 1$, 
\bea
x_1 &=& -\frac{q}{1-q^2}\, , \;\; z_1 = \frac{-2 q + q^3}{1-q^2}\, , \;\; 
y_1 = \frac{1}{1-q^2}\,, \nonumber\\
x_2 &=& q\, , \;\; y_2 = -1 \,.
\eea
For $n \geq 2$, $x_n$ and $y_n$ satisfy the recursion relations
\bea
x_n + q y_n &=& -(x_{n-1} + q y_{n-1}) - \frac{2}{n!}(q + n z_1) - 2 
\sum_{k=1}^{n-2}\, \frac{x_k}{(n-k)!}\,,\nonumber\\
q x_n + y_n &=& - (q x_{n-1} + y_{n-1}) - \frac{2}{n!}(-1 + n y_1) - 2 
\sum_{k=1}^{n-2}\, \frac{y_k}{(n-k)!}\,.
\eea         
When $q = \pm 1$, the expansion in eq.(25) reduces to $a_i \ad_j = 
\delta_{ij} \pm \ad_j a_i$ and all other terms vanish identically.

The partition function of a free $M$-level system defined in eq.(17) is (for 
$|q| < 1$)
\begin{equation}
Z(x_1,...,x_M) = \prod^{M}_{i=1} \, \frac{1}{1-x_i} + \prod^{M}_{i=1} \, (1 - x_i) - \sum_{i=1}^M x_i -1
\end{equation}
and the number of independent $N$-particle states is the sum of Bose and 
Fermi counting rules
\begin{equation}
D(M,N) = \left( \stackrel{\displaystyle M+N-1}{N} \right) + 
\left(\stackrel{\displaystyle M}{N} \right)
\;\;\; N \geq 2 \,.
\end{equation}
We point out that our simple interpolation defined by eq.(17) is equivalent to 
the statistics introduced by Wu {\it et al.} [7] and Scipioni [8]. 
The construction
of Wu {\it et al.} is based on two vacuums $|\pm \rangle$ and on the 
commutation rules containing the $g$ - operator:
\bea
a_i \ad_j - g \ad_j a_i &=& \delta_{ij} \,,\nonumber\\
g |\pm \rangle = \pm |\pm\rangle\,.
\eea
They introduced the $\phi$-vacuum as a linear combination of the 
$|\pm \rangle$ vacuums, $|\phi\rangle = cos\phi |+\rangle + sin\phi |-\rangle$
and defined the corresponding Fock representation built on $|\phi\rangle$. 
The $q$ parameter in eq.(17) is then related to the angle $\phi$ through $cos\phi = \sqrt{(1+q)/2}$. In our approach we start with one vacuum from the 
beginning and (except $a_i,\ad_i$), no additional operators appear in eq.(25).

The physical consequences of the minimal interpolation follow from the grand 
partition function, eq.(28). It consists of the bosonic and fermionic partition
 functions from which one-particle states are subtracted. Such a partition 
function mainly has a Bose character since, for a large number of particles,  
the 
symmetric (bosonic) subspace is much larger than the antisymmetric (fermionic) 
subspace.  
Therefore, the whole spectrum of bosonic phenomena can
 be found here: Bose condensation [7], black body radiation [8]. 
The effects of the antisymmetric states keep trace only in corrections to the 
ordinary Bose phenomena, disappearing completely in the high temperature limit.

It is worth mentioning that the statistics of the type described in this letter can be discussed from the point of view of possible violation of Bose 
statistics. Some 
analysis has been done [16], comparing the experimental limits on the 
$Z$ boson decay 
into two photons with the theoretical consideration based on a general 
phenomenological model of Bose symmetry violation. 
In the model of minimal mixing between bosons and fermions, the $q$-parameter 
would be $q<10^{-2}$.

For comparison, we mention that another simple interpolation between Bose and 
Fermi statistics [5]
\begin{equation}
a_i \ad_j - q \ad_j a_i = \delta_{ij}\; , \;\;\; |q| < 1\,,
\end{equation}
corresponds to infinite quon statistics, i.e. to the maximal interpolation 
in which every IRREP $\mu$ of $S_N$ 
contributes with the multiplicity $n(\mu) = K_{\mu,1^N} = dim(\mu)$. The number of 
independent $N$-particle states is $D(M,N) = M^N$.
\\
\\
{\bf Minimal interpolation between parabosons and parafermions}

Para-Bose and 
para-Fermi statistics of a given order $p \in N$ generalize the Pauli exclusion 
principle and also belong to the class of
permutation invariant statistics. 
The $N$-particle state of para-Bose (para-Fermi) statistics of order $p$ 
cannot be antisymmetrized (symmetrized) in more than $p$ indices, which means 
that the allowed Young tableaux are restricted to those with at most $p$ rows 
(columns).
They are defined through trilinear relations:
\begin{equation}
[\,(\ad_i a_j \pm a_j \ad_i), \ad_k \,] = \frac{2}{p}\,\delta_{jk}\ad_i\;,\;\;
 i,j,k = 1,2,...,M
\end{equation}
with the unique vacuum $|0\rangle$ and the following conditions: $\langle 0|0 
\rangle = 1$, $a_i|0\rangle = 0$, $a_i\ad_j|0\rangle = \delta_{ij}|0\rangle$. 
 The upper (lower) sign in eq.(32) corresponds to the para-Bose (para-Fermi) 
algebra and $p$ is an integer. 

It was shown that no interpolation between para-Bose and para-Fermi statistics 
through deformed trilinear relations was possible [6], since states of the 
negative norm appeared. However, in [14] it is suggested that such an 
interpolation is possible through a continuous family of generic matrices. 
The concrete construction was performed through deformed Green's oscillators 
obeying infinite quon statistics. The corresponding statistics belongs to the 
class of infinite statistics and is similar 
to that of Greenberg [5] and reduces to it for $p=1$ and $p=\infty$.

Here we suggest a new family of generic matrices interpolating between 
para-Bose and para-Fermi generic matrices defined by eq.(15). 
It follows from eq.(32) that the coefficients $f^{p,\epsilon}(\pi)$ for 
parastatistics, $\epsilon = +\,$ (parafermions)$/-\,$ (parabosons) of order $p$, 
satisfy recursion relations [6,11], and that
\begin{equation}
f^{p,-\epsilon}(\pi) = (-)^{I(\pi)}f^{p,\epsilon}(\pi)\,.
\end{equation}
This recursion relation implies that para-Bose and para-Fermi statistics of 
order $p$ are dual to each other (see eq.(10) and equations following eq.(11)).

The generic matrices $A_{1^N}^{p,\epsilon}$, eqs.(6),(33), are hermitian and 
if $p$ 
is a positive integer their eigenvalues are non-negative [3]. Using the 
results of
[12], we find that the eigenvalues $(\Lambda^{p,\epsilon})_{\mu}$, 
corresponding only 
to one of equivalent IRREP's $\mu$ of $S_N$, are 
\begin{equation}
\Lambda^{p,\epsilon}_{\mu} = \sum_{\pi \in S_N} f^{p,\epsilon}(\pi) \chi^
{\mu}(\pi)\,,
\end{equation}
where $\chi^{\mu}$ is the character of the IRREP $\mu$. We point out that the 
eigenvalues corresponding to all other (except one) equivalent IRREP's 
$\mu$ are identically zero.

Applying the interpolation between dual statistics, eq.(15), to para-Bose 
and para-Fermi statistics, we have
\begin{equation}
A^{p,q}_{1^N} = \frac{1+q}{2}A^{p,-}_{1^N} + \frac{1-q}{2}A^{p,+}_{1^N}\,,
\end{equation}
where $|q| \leq 1$. If $|q| < 1$, there are at most two positive eigenvalues 
corresponding to the equivalent IRREP's $\mu, (\mu \neq
1^N, N)$ of $S_N$
and if $\mu = 1^N,N$, then
\begin{equation}
\Lambda^{p,q}_{\mu} = \frac{1+q}{2}\Lambda^{p,-}_{\mu} + 
\frac{1-q}{2}\Lambda^{p,+}_{\mu}\,,
\end{equation}
where $\Lambda^{p,\epsilon}_{\mu}$ are given by eq.(34). The 
$\Lambda^{p,\epsilon}_{\mu} = 0$ if the number of rows of 
$\mu$, $l(\mu)$, is $l(\mu) > p$ for $\epsilon = -$ and 
$l(\mu^T) > p$ for $\epsilon= +$.

Note that the eigenvalues corresponding to all (except at most two) 
equivalent IRREP's $\mu$ vanish identically. Hence, at most two equivalent 
IRREP's $\mu$ ($\mu \neq 1^N,N$) contribute to eqs. (7),(9). We therefore call 
the above interpolation minimal since $n(\mu) \leq 2$.

%

The probability of finding a generic state 
$\ad_{i_1}\cdot\cdot\cdot\ad_{i_N}|0\rangle$ with mutually different indices, 
in all equivalent IRREP's $\mu$ of $S_N$, is 
\begin{equation}
w(\mu) = \frac{K_{\mu,1^N}}{N!}\,[\frac{1-q}{2}\Lambda^{p,+}_{\mu} + 
\frac{1+q}{2} \Lambda^{p,-}_{\mu}]\,,
\end{equation}
which generalizes the result for mixing of bosons and fermions. The 
relations (3)-(5) for the minimal interpolation between para-Bose and para-Fermi statistics 
of order $p$ can be obtained similarly as in [11].

Let us point out that the above minimal interpolation can be obtained by 
generalizing the statistics of Wu {\it et al.} [7] and Scipioni [8]. The 
relations in (32) become
\begin{equation}
[(\ad_i a_j + g a_j \ad_i), \ad_k] = \frac{2}{p} \delta_{j,k} \ad_i\,,
\end{equation}
where $i,j,k = 1,2,...,M$ and $g|\pm\rangle = \pm|\pm\rangle$. Choosing one
vacuum \\ $|\phi\rangle = cos\phi|
+\rangle + sin\phi|-\rangle$, one obtains the same
 statistics as in eq.(35) with $cos\phi = \sqrt{(1+q)/2}, \phi \in [0,\pi/2]$.
\\
\\
{\bf Simple interpolation between anyonic-like statistics}

Finally, let us mention that the above consideration on duality and a simple 
interpolation can be extended to anyonic-like generalized statistics, which are
not invariant under the permutation group. 
We call them anyonic-like statistics by analogy 
with the anyonic interpolation between Bose and Fermi statistics, where we
interpolate between any 
two permutation invariant dual statistics. 
 
The anyonic-like generic matrix 
$A_{1^N}^{\phi}$ can be obtained from any permutation invariant generic 
matrix $A_{1^N}$, eq.(6), in the following way:
\begin{equation}
A_{1^N}^{\phi} = D_{1^N}^{\phi}\, A_{1^N}\, D_{1^N}^{-\phi} \,,
\end{equation}
where $D^{\phi}_{1^N}$ is the $N!\times N!$ diagonal matrix with matrix 
elements,
\begin{equation}
(D^{\phi}_{1^N})_{\pi\sigma} = e^{i\phi(\pi)}\,\delta_{\pi\sigma}
\end{equation}
and $\phi(\pi)$ are real coefficients (i.e. function from $S_N$ to real 
numbers). If the generic matrix $A_{1^N}$ is hermitian, then the corresponding 
anyonic generic matrix $A^{\phi}_{1^N}$ in eq.(40) is hermitian with the same 
spectrum as $A_{1^N}$. If $A_{1^N} = \sum_{\pi\in S_N} f(\pi)
R(\pi)$, then $A^{\phi}_{1^N} = \sum_{\pi\in S_N} f(\pi) R^{\phi}(\pi)$, where 
$R^{\phi} = D^{\phi}RD^{-\phi}$, i.e. $R^{\phi}_{\alpha\beta}(\pi) = 
e^{i[\phi(\alpha)-\phi(\beta)]}R_{\alpha\beta}(\pi)$ is equivalent to 
the regular representation. 
The non-generic matrix $A^{\phi}_{\lambda}(i_1,...,i_N)$, $\lambda \neq 1^N$, 
$|\lambda|=N$, is similarly defined as $A_{\lambda}$, see the relation 
preceding eq.(6). 
However, the phases in $D^{\phi}$, eq.(40), are of a more general form 
$\phi (\pi;\lambda)$ depending on permutation $\pi$ and the multiplicities 
of equal indices. Thus, we point out that generally 
no simple redefinition of states by an insignificant phase factor 
is possible. It could be done for the generic matrix $A_{1^N}$ alone, but not
for $A^{\phi}_{\lambda}$ for all partitions $\lambda \neq 1^N$, 
$|\lambda| = N$ simultaneously. 
There is a large class of anyonic-like algebras which can be obtained by non-
linear transformation on $\ad_i$, $a_i$ from permutation invariant algebras, 
but this is not true for every anyonic-like statistics in general.

The dual generic matrix is defined by
\begin{equation}
(A^{\phi}_{1^N})^d = (A^d_{1^N})^{\phi} = D_{1^N}^{\phi}\,A_{1^N}^d\,D^{-\phi}_
{1^N} = \sum_{\pi\in S_N} f^d(\pi) R^{\phi}(\pi) \,.
\end{equation}
The anyonic-like generic matrices $A_{1^N}^{\phi}$ and $(A_{1^N}^{\phi})^d$ are 
hermitian and have the same eigenvectors and eigenvalues.

The simple interpolation between the $A_{1^N}^{\phi}$ and 
$(A_{1^N}^{\phi})^d$, defined by eq.(15), has non-negative eigenvalues 
for $|q|\leq 1$.

The simplest examples of anyonic-like statistics related to 
permutation invariant statistics by eq.(40) are obtained as special cases 
$|q_{ij}|=1$, $i,j=1,2,...,M$, of the operator algebra defined by 
$a_i\ad_j - q_{ij}\ad_j a_i = \delta_{ij}$, $q_{ij}^{\ast} = q_{ji}$, 
$i,j=1,2,...,M$, investigated in [17],[18]. They can be obtained by regular 
non-linear 
mapping from fermions and/or bosons. However, anyonic generic matrices
are not permutation invariant, although they are in a simple way related 
to Bose and Fermi generic matrices, eq.(39). 
\\
\\
{\bf Conclusion}

We have considered general permutation invariant statistics in the second 
quantized approach. Particularly we have investigated generic matrices 
$A_{1^N}$ and their dual matrices $A^d_{1^N}$. Then we have suggested a 
simple interpolation between these two types of statistics. The permutation 
invariant 
statistics considered is completely determined (including the probabilities 
of finding 
IRREP's $\mu$ of $S_N$ in all decompositions) by the functions $f(\pi)$, 
$\pi \in S_N$, for all $N$ that lead to non-negative eigenvalues of $A_{1^N}$. 

Particularly, we have presented new minimal interpolations between (para)bosons 
and (para)fermions of order $p$ and established a connection with the 
mixing of bosons and fermions proposed by Wu {\it et al.} [7] and Scipioni [8].

Finally, we have proposed an extension of our analysis to anyonic-like 
statistics, 
which are related to permutation invariant statistics by eq.(40). 
\\
Simbolically, we can write
\begin{eqnarray}
& &A_{1^N} \Longleftarrow \;\;A^q_{1^N} \;\;\;\Longrightarrow A^d_{1^N} 
\nonumber\\
& &\;\,\Downarrow \;\;\;\;\;\;\;\;\;\;\;\;\;\; \Downarrow 
\;\;\;\;\;\;\;\;\;\;\;\;\;\;\; \Downarrow \nonumber\\
& &A_{1^N}^{\phi} \Longleftarrow (A^{\phi}_{1^N})^q  \Longrightarrow (A^{\phi}_
{1^N})^d\,,
\end{eqnarray}
where the interpolating generic matrices are defined by eq.(15). 

The physical properties 
of the minimal mixing of bosons and fermions were investigated in [7],[8] and 
the properties of the minimal mixing between parabosons and parafermions, 
as well as mixing between anyons, are under investigation.
\\
{\bf Acknowledgement}

We thank M.Milekovi\'c and D.Svrtan for useful discussions.


%
%

%
%

\end{document}